\documentclass[twocolumn,prstab,floatfix]{revtex4}
\usepackage{graphicx}
\usepackage{amsmath}
\usepackage{latexsym}
\usepackage{amsfonts}
\usepackage{amssymb}

\newcommand{\md}{\mbox{d}}

\begin{document}

\title{Equilibrium ion distribution in the presence of clearing
electrodes and its influence on electron dynamics}
\author{Georg H.~Hoffstaetter}
\author{Christian Spethmann}
\affiliation{Laboratory for
Elementary Particle Physics, Cornell University, Ithaca, New York 14853}

\begin{abstract}
Here we compute the ion distribution produced by an electron beam when
ion-clearing electrodes are installed. This ion density is established
as an equilibrium between gas ionization and ion clearing. The
transverse ion distributions are shown to strongly peak in the beam's
center, producing very nonlinear forces on the electron beam. We will
analyze perturbations to the beam properties by these nonlinear
fields.  To obtain reasonable simulation speeds, we develop fast
algorithms that take advantage of adiabatic invariants and scaling
properties of Maxwell's equations and the Lorentz force.

Our results are very relevant for high current Energy Recovery Linacs,
where ions are produced relatively quickly, and where clearing gaps in
the electron beam cannot easily be used for ion elimination. The
examples in this paper therefore use parameters of the Cornell Energy
Recovery Linac project.  For simplicity we only consider the case of a
circular electron beam of changing diameter. However, we parameterize
this model to approximate non-round beams well.  We find suitable
places for clearing electrodes and compute the equilibrium ion density
and its effect on electron-emittance growth and halo development. We
find that it is not sufficient to place clearing electrodes only at
the minimum of the electron beam potential where ions are accumulated.
\end{abstract}

\maketitle

\section{Introduction}

Several processes can contribute to the production of ions in the
vicinity of accelerated electron beams \cite{Hoffstaetter05-01}. These
positively charged ions are attracted to the negative
beam. Anti-proton accumulation systems have suffered under the field
from accumulated ions \cite{Poncet93, Zhou93-01}, and so
have electron beams \cite{Clarke93, Zobov06}.

Ions can damage the beam in various ways. Firstly, they produce a
focusing field that changes in a very nonlinear way with an electron's
distance from the beam's center. The resulting nonlinear motion can
increase the beam emittance, can lead to particle loss, and can
produce a halo around the beam. Secondly, the ion distribution
oscillates within the electron beam, while the electrons are attracted
to the ion distribution. This system of coupled oscillators can become
unstable \cite{Zhou93-02, Raubenheimer95, Byrd97, Kim05} and lead to
large transverse beam oscillation amplitudes or to an increase of the
apparent transverse beam size.

It is therefore important to reduce the density of ions in the
vicinity of the beam to a tolerable amount. Different accelerators
achieve this differently. Storage rings typically use ion-clearing
gaps. These are short gaps in the filling pattern that lead to an
absence of focusing forces for the ions every time this gap has
traveled around the ring. When the gap length and frequency are
chosen suitably, the ions get over-focused which lets the ions
oscillate with increasing amplitude until they have moved outside the
beam region. The length of the beam-filled region is typically a few
microseconds long for a large accelerator, whereas the gap is
typically shorter than one microsecond. In pulsed linacs, the gaps are
often much longer and allow ions to drift out of the beam region.

In rings with coasting beams, there is no ion-clearing gap, and
obviously the beam cannot be turned off regularly. Similarly, in
Energy Recovery Linacs (ERLs) \cite{Hoffstaetter05-01} where the beam's
energy is dumped in RF cavities and is immediately used to accelerate
new electrons, one cannot easily turn off the beam (because this would
interrupt the ERL process) and one can also not easily introduce short
gaps in the beam (because this would disrupt the gun or the linac that
injects large currents into the ERL). In both of these cases,
ion-clearing electrodes may have to be used \cite{Bulyak93, Bulyak96}.

The electron beam diameter varies along the accelerator, and this variation
produces longitudinal forces, guiding ions to a location where the
electron density is relatively large, typically close to the waists of
the electron beam. Clearing electrodes are placed along the beam-line
at such places of ion accumulation.

Because these longitudinal forces are relatively weak, it typically
takes a few milliseconds for an ion to move from the place where it is
produced to an ion-clearing electrode a few meters down the
beam-line. During this time, new ions are created, leading to a
remnant equilibrium ion density that establishes itself in the
presence of clearing electrodes. Creating a number of ions per length
as large as the number of electrons per length typically takes only a
few seconds, even for very good vacuum in the nTorr range
\cite{Hoffstaetter05-01}. Because the motion to clearing electrodes 
that are spaced
many meters apart takes several milliseconds, such electrodes tend to
produce a linear ion density that is in the order of about one part in
a thousand of the linear charge density in the beam.

Computing an equilibrium density is typically very time
consuming. Here we show how to use scaling properties of the Maxwell
equations and the Lorentz force, as well as the adiabatic invariance
of the action integral, to compute the equilibrium ion density very
efficiently.

We subsequently apply this technique to analyze the Cornell ERL
project \cite{Hoffstaetter05-02}. We investigate whether the remnant
ion density after placement of clearing electrodes can lead to any of
the discussed damages to the electron beam, so that additional clearing
electrodes would have to be installed. We assume the parameters listed
in Tab.~\ref{tb:erlpara}, where the ionization cross-section is taken
from \cite{Rieke72} for 5GeV. To be conservative we used the high energy
cross section for all energies, even though for lower energies the
cross-section is up to 40\% smaller. The gas density of
$3\cdot 10^{13}$m$^{-3}$ corresponds to a pressure of 1nTorr for room
temperature sections of the accelerator. The unnormalized emittances
are obtained with the relativistic $\gamma$ by
$\epsilon_x=\varepsilon_{nx}/\gamma$.
\begin{table}
\caption{Parameters of the Cornell ERL used for the examples in this
paper.\label{tb:erlpara}}
\begin{tabular}{l|l|l}
\hline
Normalized emittances & $\varepsilon_{nx} = \varepsilon_{ny}$ & $0.3\cdot 10^{-6}$m      \\
Energy spread                & $\sigma_\delta$            & $2\cdot 10^{-4}$         \\
Electron current             & $I$                       & $0.1$A                   \\
Bunch charge                 & $Q$                       & 77pC                     \\
Injected energy              & $E_{{\rm in}}$            & 10MeV                    \\
Top energy                   & $E_{{\rm top}}$           & 5GeV                     \\
Dominant ion abundance       & $H^+$                     & 98\%                     \\
Ionization cross-section     & $\sigma_{{\rm col}}$      & $3.8\cdot 10^{-23}$m$^2$ \\
Gas density for warm sections& $p_{{\rm line}}$          & $3\cdot 10^{13}$m$^{-3}$ \\
Gas density for cryogenic linac& $p_{{\rm LINAC}}$       & $3\cdot 10^{11}$m$^{-3}$ \\
\hline
\end{tabular}
\end{table}

\section{Computation of the ion equilibrium}

\subsection{The 3D beam-beam force}

For a rotationally symmetric electron beam, the density is given by
\begin{equation}
\rho_e(r,s) = \lambda\frac{1}{2\pi\sigma(s)^2}e^{-\frac{r^2}{2\sigma(s)^2}}\ .
\end{equation}
where $\lambda$ is the linear particle density. With Gauss' law, the
transverse velocity kick on a singly charged,
non relativistic ion becomes
\begin{equation}
\Delta v_r = - \; \frac{2Ncr_p}{Ar} \; 
\left( 1 - \exp \left[ \frac{-r^2}{2 \sigma^2} \right] \right)\ ,
\label{eq:gauss_kick}
\end{equation}
where $\sigma$ is the rms width of the electron-beam profile, $r$ is
the distance from the beam centerline, $N$ the number of electrons in
the bunch, $r_p$ the classical proton radius and $A$ the atomic weight
of the ion.  For non-round Gaussian electron beams, the transverse
kick of the ion can be evaluated analytically in terms of complex
error functions \cite{Talman_87}.

\begin{figure} 
\includegraphics[width=0.4\textwidth]{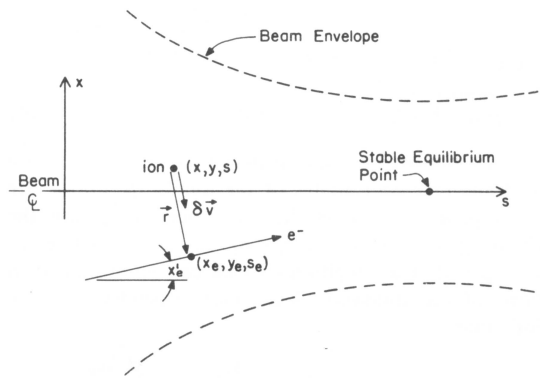}
\caption{Impulse transmitted on an ion
from a passing beam electron \cite{Sagan_91}}
\label{fg:force-fig}
\end{figure}

\begin{figure}
\includegraphics[width=0.5\textwidth]{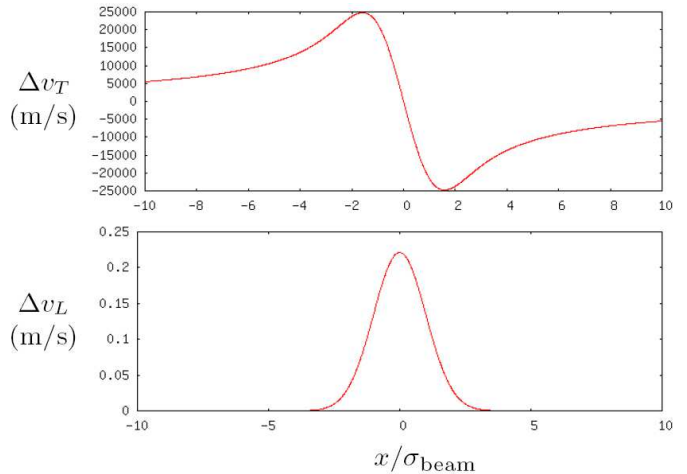}
\caption{Transverse (top) and longitudinal (bottom) kick component in
  m/s vs.~transverse ion position in units of the rms beam size.}
\label{fg:gauss_force} 
\end{figure}

If the size of the electron beam changes as a function of $s$,
the ion will also experience a longitudinal force from the passing
bunch as shown in Fig.~\ref{fg:force-fig}. 
Linearizing in the slopes of the electron trajectories, this longitudinal
force can be expressed as a function of transverse force components
and beam parameters \cite{Sagan_91}:
\begin{equation}
\Delta v_s = \left[ - \alpha_x \varepsilon_x + (\eta \tilde{\sigma}_\delta)
(\eta' \tilde{\sigma}_\delta) \right] \frac{\partial \: \Delta v_x}
{\partial x} - \alpha_y \varepsilon_y \frac{\partial \: \Delta v_y}
{\partial y}\ .
\end{equation}
The velocity changes that are produced by one bunch passage can be seen
in Fig.~\ref{fg:gauss_force}. Here we used the values of
Tab.~\ref{tb:erlpara} for top energy and Twiss parameters of
$\beta_x=\beta_y=2$m, $\alpha_x=\alpha_y=1$.

In our restriction to rotationally symmetric beams, we have a
dispersion of $\eta=0$, $\alpha=\alpha_x=\alpha_y$,
$\varepsilon=\varepsilon_x=\varepsilon_y$, and the longitudinal kick
becomes
\begin{equation}
\Delta v_s = - \alpha \varepsilon \left( \frac{\partial \: \Delta v_x}
{\partial x} - \frac{\partial \: \Delta v_y}
{\partial y} \right)\ .
\end{equation}
For the beam parameters we studied, the longitudinal kick is typically
five orders of magnitude weaker than the transverse kick, as shown in
Fig.~\ref{fg:gauss_force}. In the special case of rotational symmetry,
the longitudinal kick on the ion will be in the same direction for all
positions. This is not true in the general case; if $\alpha_x$ and
$\alpha_y$ have different signs, the kick direction can depend on the
distance from the centerline \cite{Sagan_91}.

\subsection{Adiabatic invariants of ion motion}

In a brute force simulation, the state of each ion is characterized 
by its x and s position, as well as by the velocities in both directions.
Every kick has to be calculated separately from the beam-beam force,
which is assumed to be a function of the x and s position. 
Such a simulation would be very inefficient because there are two
different timescales: A large number of kicks 
is required to resolve the sharply focused transverse ion motion,
but the longitudinal motion hardly changes in a single oscillation.
It is therefore not possible to simulate the motion of a realistic 
number of ions in this way. 

The period of transverse oscillations in the beam's potential
\cite{Groebner75, Regenstreif76, Bosi82} depends on the average beam
density, and for typical high-brightness beam parameters as those of
Tab.~\ref{tb:erlpara} it is of the order of several to tens of bunch
crossings. During one transverse oscillation 
the longitudinal position $s$ of the ion
changes typically only by a fraction of a millimeter even at the largest
possible ion speeds.

Ions thus oscillate in a potential that slowly changes over many
periods.  Because of this, an ion's action integral over one
oscillation period
\begin{equation}
J=m\oint v_x \; \mbox{d} x 
\label{eq:sint}
\end{equation}
is an adiabatic invariant of the motion. This can be used to
drastically speed up the simulation.

In this improved simulation, the state of the ion is now described by
its longitudinal position $s$, its longitudinal speed $v_s$ and its
action integral $J$. For each position $s$, $J$ is a measure of the
transverse oscillation amplitude $a$, because the action increases
with $a$. This can most easily be seen in $(x,p_x)$ phase space where
the graph $(x(t),p_x(t))$ for any oscillation is a closed curve, and
$J$ is the enclosed area.  A closed curve that starts with a larger
oscillation amplitude has to completely enclose one that starts with a
smaller amplitude because different phase-space trajectories cannot
cross.  Solving the motion in $(J,s)$ rather than in $(x,s)$
coordinates has two advantages: (1) The degrees of freedom are reduced
from 2 to 1. (2) Because $\Delta v_s$ changes much slower than $\Delta
v_x$, the integration steps can be vastly increased, typically by
about a factor of 10000. If the density at $s$ is to be computed, it
is not enough to know the action at $s$, but one oscillation is
sufficient to compute the density contribution of particles with
action J.

The longitudinal acceleration averaged over one oscillation is given
by the time average over individual kicks
\begin{eqnarray}
\label{eq:vdot}
\dot{v}_s = \langle \Delta v_s \rangle = - \alpha \varepsilon \; 
\left<   \frac{\partial \: \Delta v_x} {\partial x}  
+ \frac{\partial \: \Delta v_y}{\partial y} \right>
\nonumber
\end{eqnarray} 
where
\begin{equation}
\langle \Delta v_s \rangle = \frac1T \; \sum_{\mbox{\tiny osc}}
\Delta v_s\ .
\end{equation} 
Calculating this time average again and again for each ion is
computationally demanding and wasteful. Instead, we pre-compute a table
of possible values of this time average for many beam sizes $\sigma$
and amplitudes $a$ of the oscillation.

The program was further accelerated by noting that the time averaged
kick as a function of $J$ can be calculated for a typical standard
beam size and then rescaled for regions with other beam sizes.

To derive this scaling behavior, we use a general scaling theorem for
Maxwell's equations and the Lorentz force. If $\vec E(\vec x,t)$,
$\vec B(\vec x,t)$, $\rho(\vec x,t)$, and $\vec j(\vec x,t)$ satisfy
Maxwell's equation, and a charged particle with mass $m$ and
trajectory $\vec r(t)$ satisfies the Lorentz force equation, then the
following scaled quantities also satisfy these equations,
\begin{align*}
\vec E_s(\vec{x},t)  & = \alpha^2 \; \vec E(\alpha \vec{x}, \alpha t) \\
\vec B_s(\vec{x},t)  & = \alpha^2 \; \vec B(\alpha \vec{x}, \alpha t) \\
\rho_s(\vec{x},t)    & = \alpha^3 \rho (\alpha \vec{x}, \alpha t) \\
\vec{j_s}(\vec{x},t) & = \alpha^3 \vec{j} (\alpha \vec{x}, \alpha t)\\
\vec r_s(t) & = \alpha^{-1} \vec r(\alpha t) \\
m_s & = \alpha m\ .
\end{align*}
The effect of this scaling transformation is to reduce spatial
distances and time intervals by a factor $\alpha$, so that velocities
remain unchanged. Also, the charge and current densities are increased
by a factor $\alpha^3$, so that the total charge in a scaled volume
remains constant.

We now want to find out how the spatial derivative of the transverse
kicks changes when the transverse scale of the beam is reduced by a
factor $\alpha$. If the rotationally symmetric beam produces a
field $\vec E(x,s)$, the 3-D scaled field would be $\alpha^2\vec
E(\alpha x,\alpha s)$. However, only the transverse scale changes from
one beam width to another, and the longitudinal scale stays the same,
signified by the unchanged line density. The charge within a 3-D
scaled volume is therefore smaller by $1/\alpha$, and the field after
transverse scaling is therefore $\vec E_{1/\alpha}(x,s) = \alpha \vec
E(\alpha x,s)$.

The transverse scaling transformation therefore gives
\begin{align*}
\left( \frac{\partial \; \Delta v_x}{\partial x} \right)_{1/\alpha}
& = \left( \frac{q}{m} \right)_{\mbox{\tiny ion}} \;
\frac{\partial}{\partial x} \; \int \vec{E}_{1/\alpha} (x, ct) 
\mbox{ d} t \\
& = \left( \frac{q}{m} \right)_{\mbox{\tiny ion}} \;
\frac{\partial}{\partial x} \; \frac{1}{c}\int \alpha \;
\vec{E} (\alpha x, s) \mbox{ d} s \\
& = \alpha^2 \; 
\left( \frac{q}{m} \right)_{\mbox{\tiny ion}} \;
\frac{\partial}{\partial \tilde{x}} \;
\int  
\vec{E} (\tilde{x}, ct) \mbox{ d} t \\
& = \alpha^2 \; \left( \frac{\partial \; \Delta v_x}{\partial x} \right) .
\end{align*}
>From this, we see that a rescaling of the two relevant length 
scales $\sigma$ and $a$ has the effect
\begin{equation} 
\frac{\partial \; \Delta v_x}{\partial x} \; \left( \frac{a}{\alpha}, 
\frac{\sigma}{\alpha} \right)
= \alpha^2 \; \frac{\partial \; \Delta v_x}{\partial x} 
\; (a, \sigma)\ .
\end{equation}
This is true for both derivatives, so that the time average of 
their sum multiplied by $\sigma^2$ is only a function of $a/\sigma$,
\begin{equation} \sigma^2 \; \left< \frac{\partial \: \Delta v_x}
{\partial x} + \frac{\partial \: \Delta v_y}
{\partial y} \right>  
= g(a/\sigma)\ .
\end{equation}
Our goal is to express this time average as a function of the beam
size $\sigma$ and the action integral $J$, so we also have to find the
functional dependence of $J$ on $a$ and $\sigma$.  As noted above, the
electric field at a location where the beam size is reduced by a
factor $\alpha$, is scaled in the transverse to $\alpha E(\alpha
x,s)$.  To obtain scaled transverse ion motion, we do not change the
mass of the ions, i.e. $m_s=m$. The transverse velocity is then not
changed by transverse scaling, and according to Eq.~(\ref{eq:sint}),
the action integral is reduced by a factor $\alpha$. From this we
conclude
\begin{equation}
J(a, \sigma) = \alpha \; J \left( \frac{a}{\alpha},
\frac{\sigma}{\alpha} \right)\ .
\label{eq:jscale}
\end{equation}
Combining this result with the above analysis, we finally find
\begin{equation} 
\left< \frac{\partial \: \Delta v_x}
{\partial x} + \frac{\partial \: \Delta v_y}
{\partial y} \right>  
= \frac1{\sigma^2} \; f(J/\sigma)\ .
\end{equation}
To speed up the program, a lookup table for the function $f$ is
therefore computed for a standard beam size $\sigma$ and a range of
action integrals $J$.  The longitudinal acceleration of the ions is
then given by
\begin{equation}
\dot{v}_s = - \frac{\alpha \varepsilon}{\sigma^2} \; f(J/\sigma)\ .
\end{equation}

\subsection{Fast ion propagation to equilibrium}

The toy model for our ion simulations consists of a field-free beam region
of variable length $L$, so that the $\beta$-function is given by
\begin{equation}
\beta(s) = \beta^*[1+(\frac{s}{\beta^*})^2]
\label{eq:beta*}
\end{equation}
with waist at $s=0$. We then simulate the creation of ions
in the beampipe through ionization processes. The ionization rate
in a given volume $V$ is given by
\begin{equation} 
\frac{\Delta N}{\Delta t \Delta V} = \rho_e \rho_g \sigma_{{\rm col}} c
\end{equation}
where $\rho_e$ is the local particle density of the electrons as given
by the Gaussian beam profile $\rho_g$. By integrating over the
transverse directions, one obtains an equation for the number of ions
created in a length $L$ of the beam
\begin{equation}
\frac{\Delta N}{\Delta t \Delta L} = 
\frac{I_{{\rm beam}}}{e} \rho_g \sigma_{{\rm col}}\ . 
\end{equation}
>From this it is clear that ions are created with equal rate 
at all locations along the beam. Under the influence of the longitudinal
beam force, the ions will then slowly propagate to the point with
minimal potential.

In our simulation, we assume that clearing electrodes have been placed
at the waist, so that ions crossing the $s=0$ line are removed.
An equilibrium situation is reached when an equal number of ions are
produced and removed during a given time interval. The time needed to
reach this equilibrium depends on the length of the beam pipe and
the steepness of the longitudinal potential. Typical values in our
simulation are $10^6$ to $10^8$ bunch separations of 0.77 ns.

\begin{figure}
\includegraphics[width=0.5\textwidth]{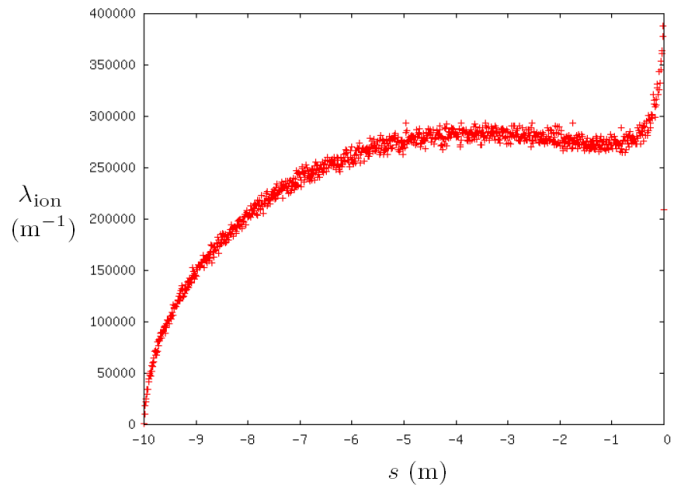}
\caption{Longitudinal ion distribution in the steady state for the 
20m beam model.}
\label{fg:20m-longdistr} 
\end{figure}

As a first example, we show a $L$=20m long part of the beam with a
waist of $\beta^*=1$m at its center. The longitudinal equilibrium ion
distribution for 10m up to the waist is shown in
Fig.~\ref{fg:20m-longdistr}. There is a sharp increase in the ion
density near the electrode at $s=0$m that is caused by two
effects. The first is that ions which have been created at all points
of the beam pass through this region and contribute to the
density. The second effect is that the $\alpha$-function becomes
smaller near the waist, so that ions which are created there
experience a weaker longitudinal force and move more slowly to the
electrode.

\begin{figure}
\includegraphics[width=0.5\textwidth]{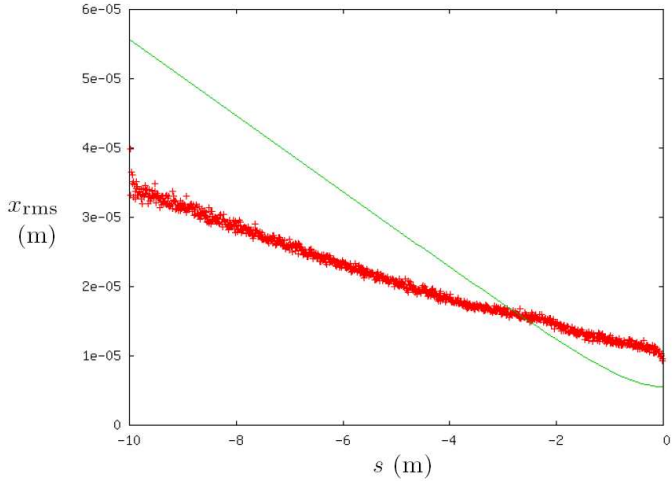}
\caption{Beam (solid line) and ion (+) transverse standard deviations as
a function of longitudinal position in the 20m beam model.}
\label{fg:20m-rms}
\end{figure}

The transverse ion distribution can be characterized by its standard
deviation, which is plotted in Fig.~\ref{fg:20m-rms}. 
Two features of this graph can be explained easily. (a) The rms width
of the ions at the start of the section (at $s=-10$m) is smaller than
that of the electron beam, even though ions are created in proportion
to the Gaussian electron distribution. The reason is that ions start
to oscillate within the electron beam after their creation and
therefore contribute to the density around zero, no matter where they
were created. (b) The ion distribution contracts less than the
electron beam. This can be explained by the adiabatic invariance and
the scaling of the action integral. Ions that are created with an
action $J$ in regions where the electron beam is wide will still have
that action by the time they reach a part of the beam that is narrower
by a factor $\alpha$. At that position, their oscillation amplitude
will have increased relative to the beam size, because the action for
a proportionally reduced oscillation amplitude would be reduced by a
factor $\alpha$ according to Eq.~(\ref{eq:jscale}).

\begin{figure}
\includegraphics[width=0.5\textwidth]{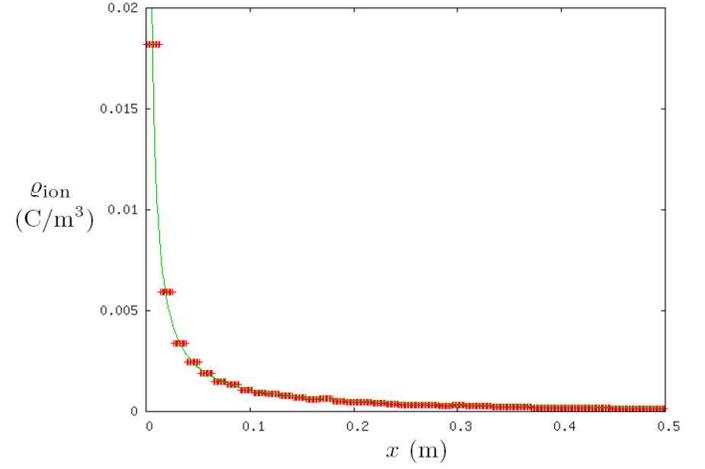}
\caption{Ion transverse density profile
at $s=\pm 1$m in the 20m beam model and comparison with a $1/r$ 
distribution.}
\label{fg:20m-density}
\end{figure}

The density profile of the transverse ion distribution at $s=\pm 1$
near the clearing electrode is shown in Fig.~\ref{fg:20m-density}.
Near the beam axis, the density diverges approximately as $1/r$, so
that it is not appropriate to fit the ion distribution to a Gaussian.

For a rotationally symmetric particle density, the number of ions
created in a length d$s$ and in a radius element d$r$ per time is
\begin{equation}
\dot n(r,s) \mbox{ d}r \mbox{d}s = c \sigma_{{\rm col}}
\rho_g \rho_e(r,s) \; 2\pi r \mbox{ d} r  \mbox{d} s .
\end{equation}

Subsequent to the creation of an ion at ($r_0, s_0$), it travels along
the beamline to $s=S(t;r_0,s_0)$ and oscillates through the electron
beam to $R(s;r_0,s_0)$. Ions created between $s_m$ and $s$ can
contribute to the distribution at $s$, and a ring element d$r$d$s$ with radius
$r$ therefore contains d$n$ ions,
\begin{eqnarray}
dn(r,s) &=&
\int_{s_m}^s
\int_0^\infty
\int_{-\infty}^{\infty}\dot n(r_0,s_0) \mbox{ d}t  
\mbox{d}r_0 \mbox{d}s_0\nonumber\\
&&\times\delta(r-R(s;r_0,s_0))\mbox{ d}r\nonumber\\
&&\times\delta(s-S(t;r_0,s_0))\mbox{ d}s\nonumber\\
&=&
\int_{s_m}^s
\frac{\dot n(r_0,s_0)}{|\partial_{r_0}R(s;r_0,s_0)|}
\frac{\mbox{d}s_0 \mbox{d}r \mbox{d}s}{|\dot S(t;r_0,s_0)|}\ ,\nonumber
\end{eqnarray}
where $r_0$ is the radius at $s_0$ that leads to $r$ when the ion
arrives at $s$ and $t$ is the time at which it arrives,
i.e. $r_0(s_0;r,s)$ and $t(s_0;r,s)$,
\begin{eqnarray}
\frac{\mbox{d}n}{\mbox{d}r \mbox{d}s}(r,s) &=&
\int_{s_m}^s
\frac{\dot n(r_0,s_0)}{|\dot S(t;r_0,s_0)|}
\left|\frac{\partial r_0}{\partial r}\right|
\mbox{ d}s_0\ .
\end{eqnarray}

The number of particles per radius interval 
$\frac{{\rm d}n}{{\rm d}r {\rm d}s}(r)$ is
clearly not zero for $r=0$, and the density close to zero
therefore has a singularity of the form
\begin{equation}
\rho_e(r) = \frac{1}{2\pi r}\frac{\mbox{d}n}{\mbox{d}r \mbox{d}s}(0)\ .
\end{equation}

Because of repulsive forces between the ions, the density at the
origin cannot diverge, and we therefore estimate the radius $r_m$
below which these forces become larger than the forces from the
electron beam. It will turn out that this radius is a very small
fraction of the electron beam's width, so that deviation from the
$1/r$ density can be neglected.

We assume that the region with radius $r_i$ in which the ion density
can be approximated as $\rho_{{\rm ion}}=\frac{A}{r}$ contains most of the
charge. At some radius, the density has to fall off faster than $1/r$
to integrate to a finite value. Our assumption therefore leads to an
overestimate of the ion force. Integrating $\rho_{\rm ion}$ up to $r_i$
leads to the ion line density $2\pi A r_i=\lambda_{{\rm ion}}$.

The region that contains most of the charge has a
radius $r_i$ that is usually smaller than the rms width of the electron
beam. Only if ions were created in a very wide region of the beam, and
it were to be focused down to an extremely small radius, could the ion
width be much larger than the electron width. However, such extreme
focusing does not occur in most beams, especially not in the Cornell
ERL.

Gauss' law within $r_i$ leads to the constant radial electric field
\begin{equation}
E_{{\rm ion}}=\frac{\lambda_{{\rm ion}}}{2\pi\epsilon_0 r_i}\ .
\label{eq:eion}
\end{equation}
The radial electric field of the electron beam has already been used in
Eq.~(\ref{eq:gauss_kick}) and for radii much smaller than $\sigma$
simplifies to
$E_e=\frac{\lambda r}{4\pi\epsilon_0\sigma^2}$. We can now find the
radius $R$ for which the ion field is as large as the electron field,
$R=\sigma\frac{\lambda_{{\rm ion}}}{\lambda}\frac{2\sigma}{r_i}$. As long as
the number of ions per length is much smaller than the number of
electrons per length, the $1/r$ scaling of the ion density is
therefore correct and not altered by the ion forces in by far the
largest part of the beam.

\section{Electron motion in the ion potential}

\subsection{Electron propagation}

\begin{figure}
\includegraphics[width=0.5\textwidth]{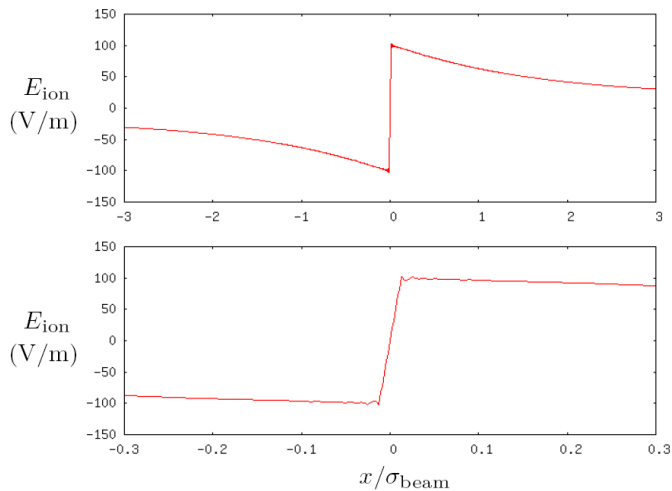}
\caption{Electric field produced by the ions 
at $s=\pm 1$m in the 20m beam model.}
\label{fg:20m-field}
\end{figure}
Because of the $1/r$ ion density characteristics, the electric field
of the ion distribution approaches a constant for small distances, as
specified in Eq.~(\ref{eq:eion}). This is shown in
Fig.~\ref{fg:20m-field}. To investigate the strength of the resulting
beam distortion, we simulated the propagation of a representative set
of beam electrons through the ion field. Our simulation was set up to
include a piece of the beampipe of length $L$, with the clearing
electrode and the waist of the beam placed in the center.

The initial electron distribution is determined by random 
betatron amplitudes $J_x$ and phases $\phi_x$ with the 
density distribution
\begin{equation}
\rho(J_x, \phi_x) = \frac{1}{2 \pi \varepsilon} \; 
e^{-J_x/\varepsilon}\ .
\end{equation}

We also investigated the effect of the beam transversing the region of
length $L$ several times, which could lead to the build-up of damaging
resonances. For this study, the linear optics has to be made periodic 
after crossing the field-free region. This can be
accomplished by including a thin half-quad with the transfer matrix
\begin{equation}
M_{q} = 
\left( \! \begin{array}{cc}
  1 & 0 \\ -kl/2 & 1 
\end{array} \right)
\end{equation}
before and after each run through the ion field. The
\mbox{$\alpha$-function} is $\alpha_-=0$ before and $\alpha_+ = \beta
\; kl/2$ after the thin quadrupole.  Simultaneously, this
$\alpha$-function must correspond to that of a drift, i.e. $\alpha_+ =
\frac{L}{2\beta^*}$.  Therefore,
\begin{equation}
kl/2 = \frac{L}{2 \beta \beta^*} = 
\frac{L/2}{(\beta^*)^2 + (L/2)^2}\ .
\end{equation} 

The simulation of the electron motion is implemented with a 
dynamical time step proportional to the distance from
the beam axis, so that the effects of the strong field near the
centerline are accounted for accurately.

\begin{figure}
\includegraphics[width=0.5\textwidth]{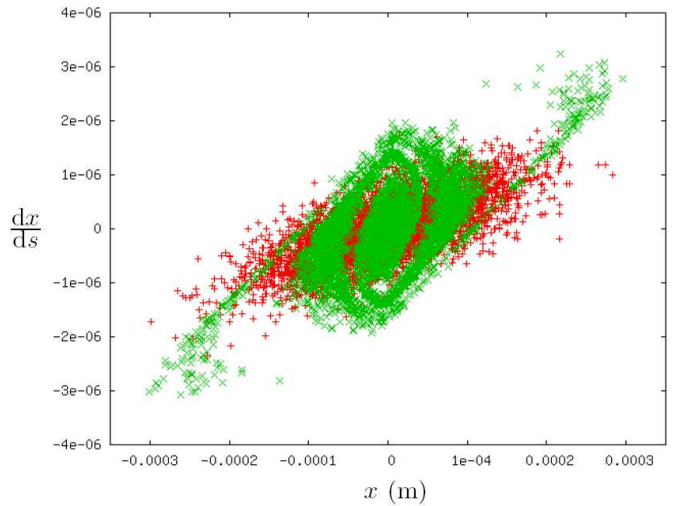}
\caption{Electron beam phase-space distribution 
after transversing the 200m ion field with $\beta^*=100$m at its center.
Dark-red $+$: phase space in a free drift,
Light-green $\times$: phase space for motion through the ion field.}
\label{fg:200m-phase}
\end{figure}

\begin{figure}
\includegraphics[width=0.5\textwidth]{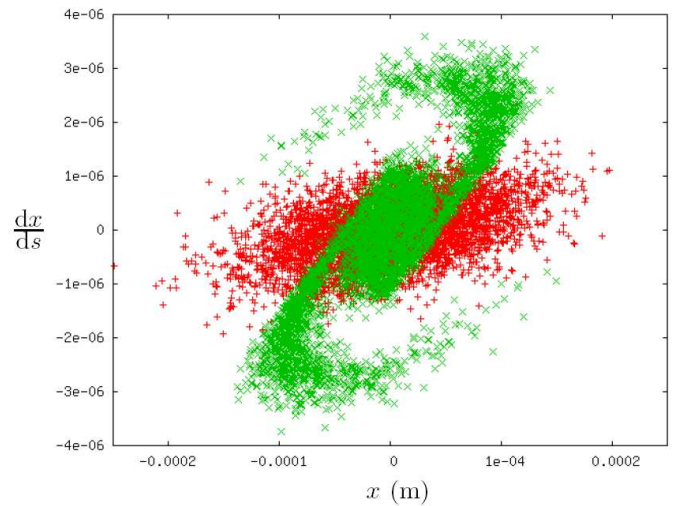}
\caption{Electron beam phase-space distribution 
after transversing the 100m ion field with $\beta^*=100$m at its center.
Dark-red $+$: phase space in a free drift,
Light-green $\times$: phase space for motion through the ion field.}
\label{fg:100m-phase}
\end{figure}

As a worst case scenario, we simulated two drift regions with 
200m and 100m length and a beam waist of $\beta^*=100$m at the 
center in both cases. The final phase-space distributions with 
and without the ion field are shown in
Figs.~\ref{fg:200m-phase} and \ref{fg:100m-phase}.

\begin{figure}
\includegraphics[width=0.5\textwidth]{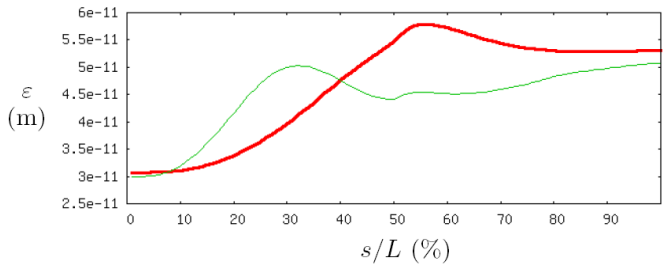}
\caption{Emittance increase while transversing the 100 m
(thick-dark-red) and 200m (thin-light-green) ion fields (horizontal
axis in percent of distance $L$).}
\label{fg:emittance}
\end{figure}

While passing through the ion field once, the emittance increases from
30.6pm to 53.1pm in the 100m field and to 50.7pm in the 200m
field, as can be seen in Fig.~\ref{fg:emittance}.  The electronic
version of this paper contains an animation of the beam as it
transverses the ion fields.

\subsection{Ion-force driven emittance growth for the Cornell ERL}

The previous two examples show that sections of only a few times 10
meters between clearing electrodes can produce intolerable emittance
growth if the beta function is large, because in that case 
the beam is not very divergent and the longitudinal force that 
clears ions is consequently small. It therefore has to be 
tested whether the optics in the Cornell
ERL provides fast enough ion motion to clearing electrodes, so that
emittance growth is limited.

\begin{figure}
\includegraphics[width=0.5\textwidth]{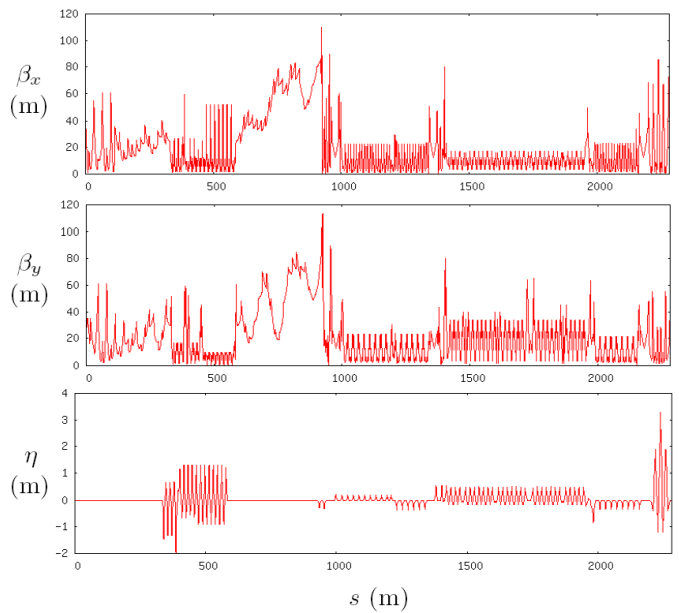}
\caption{Beam optics for the Cornell ERL. Top to bottom: horizontal $\beta$-
function, vertical $\beta$-function and horizontal dispersion.}
\label{fg:optics}
\end{figure}

\begin{figure}
\includegraphics[width=0.5\textwidth]{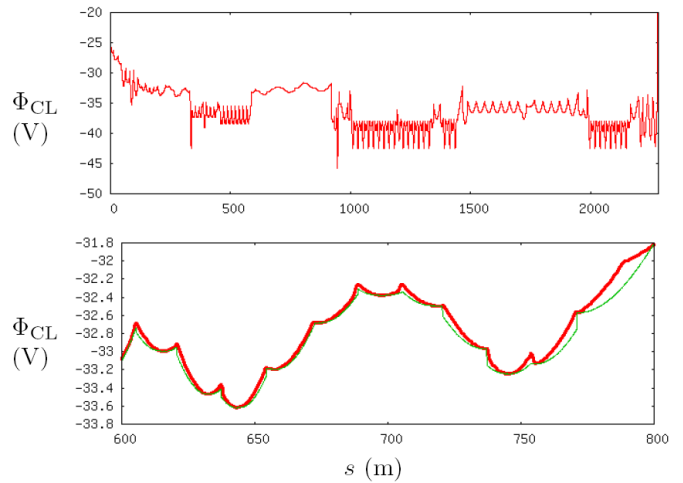}
\caption{Top: Approximate longitudinal beam potential for the Cornell
ERL. Bottom: A section of the ERL illustrating the approximation of the beam's
potential (thick-dark-red) by a round-beam model (thin-light-green).}
\label{fg:ERL_V}
\end{figure}

Because ions travel to the minima of the electrostatic potential, a
clearing electrode has to be located at every such minimum. We
therefore first calculate the potential in the beam's center with the
approximate equation \cite{Regenstreif76}
\begin{equation} 
\Phi = - \frac{I}{2 \pi \epsilon_0 c} \; 
\left[ \frac12 + \log \frac{R \sqrt{\gamma}}
{\sqrt{\varepsilon_{nx} \beta_x + (\eta_x \sigma_\delta)^2
  }+\sqrt{\varepsilon_{ny} \beta_y}} \right]\ , 
\end{equation}
for a round beam pipe with radius $R$ and a beam with the relativistic
factor $\gamma$. This equation assumes a uniform transverse density
distribution. Fig.~\ref{fg:optics} shows the relevant beam optics for
the Cornell ERL and Fig.~\ref{fg:ERL_V} shows the resulting potential
in the center of the beam.

\begin{figure}
\includegraphics[width=0.5\textwidth]{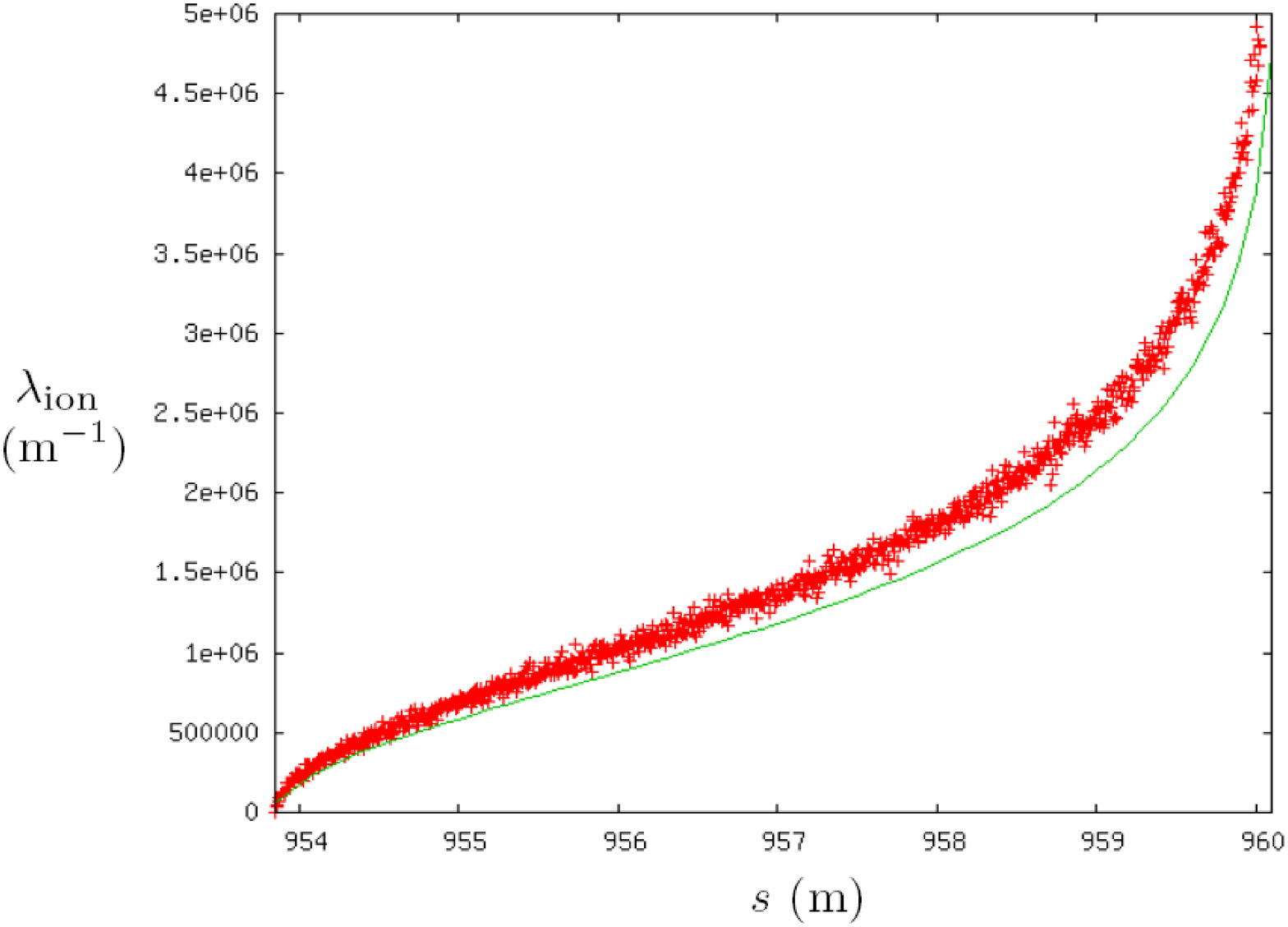}
\caption{Simulated and analytically approximated linear ion
densities near s=960m in the Cornell ERL.}
\label{fg:ERL_B}
\end{figure}

\begin{figure}
\includegraphics[width=0.5\textwidth]{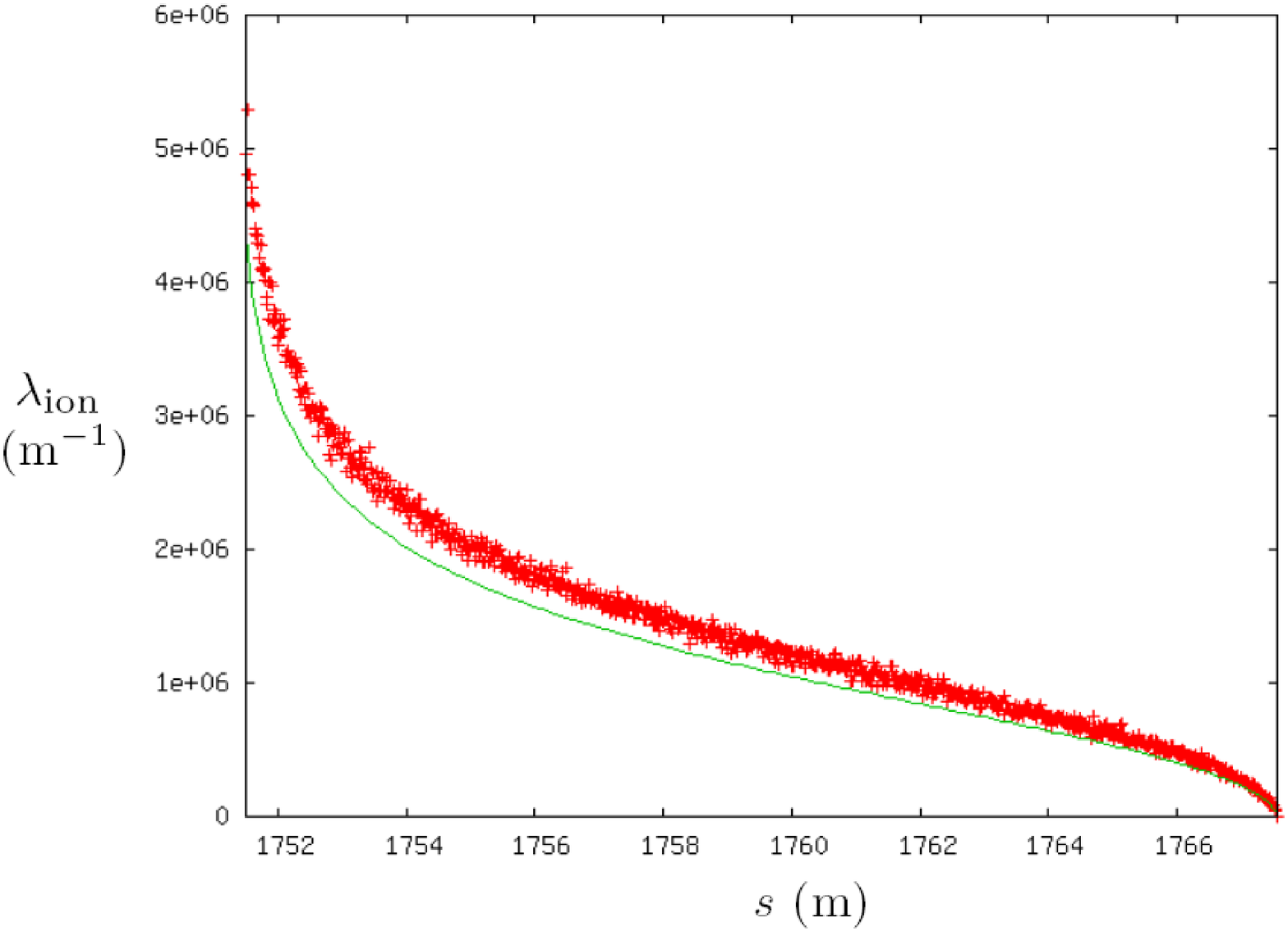}
\caption{Simulated and analytically approximated linear ion
densities near s=1760m in the Cornell ERL.}
\label{fg:ERL_A}
\end{figure}

To find a round beam approximation suitable to our simulations, we
first look at the minima of this potential and calculate the
$\beta$-function that would give the same potential at those locations
in the case of a round beam,
\begin{equation}
\beta^* = \frac{R^2 \gamma}{4 \varepsilon_n \exp \left[ 
4 \Phi \pi \epsilon_0 c / I - 1 \right]}\ .
\end{equation}
We can then define a round beam approximation for the entire ERL
lattice by using a free drift from one potential maximum to the next
with $\beta(s)$ from Eq.~(\ref{eq:beta*}).  The longitudinal ion
velocities near the center of the beam (where most of the ions are
located) depend only on this potential gradient. As will be
demonstrated by a comparison with particle tracking in
Figs.~\ref{fg:ERL_B} and \ref{fg:ERL_A}, ions that are not in the
center have the same velocity to a good approximation.  This clearing
speed determines the linear ion density also for non-round beams, and
our round beam model should therefore be a good approximation.

The number of ions created per unit length at the longitudinal position
$s_0$ during a time interval $t-t_0$ is given by
\begin{equation}
\frac{\md N_{t-t_0}}{\md s_0} = 
(t-t_0) \, \left. \frac{\md n}{\md s \, \md t} \right|_{s_0}\ ,
\end{equation}
with
\begin{equation}
\left. \frac{\md n}{\md s \, \md t} \right|_{s_0} = 
\frac{\sigma I}{e} \, \rho_n(s_0)\ ,
\end{equation}
where the number density of the residual gas is $\rho_n$, the ionization 
cross section is $\sigma$ and $e$ is the elementary charge. 
In the equilibrium state this is the same as the rate of ions 
per unit length originating from $s_0$ passing through every surface 
perpendicular to the beam axis, provided that the surface is located 
downstream from $s_0$. 

\begin{figure}
\includegraphics[width=0.5\textwidth]{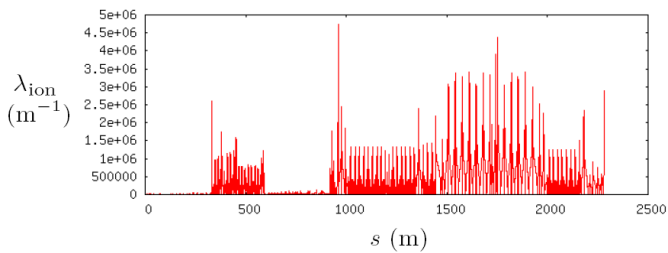}
\caption{Estimate of the linear ion density at the Cornell ERL for a round beam
approximation and clearing electrodes at the minima of the linear potential.}
\label{fg:ERL_lambda}
\end{figure}

During a time interval $t-t_0$, the total number of ions per
unit interval from $s_0$ crossing the surface is given by 
\begin{equation}
N_{s_0}(t) = (t-t_0) \; \rho_{s_0}(s) \; v_{s_0}(s)\ ,
\end{equation}
where $\rho_{s_0}(s)$ is the local linear density of ions from $s_0$
per unit length at the point $s$.  Solving for this quantity, we
obtain
\begin{equation}
\rho_{s_0}(s) = \frac{1}{v_{s_0}(s)} \,
\left. \frac{\md n}{\md s \, \md t} \right|_{s_0}\ ,
\end{equation}
where the velocity of ions from $s_0$ at $s$ is given by
\begin{equation}
v_{s_0}(s)=\sqrt{\frac{2e}{m_{\mbox{\scriptsize ion}}} \left[
\rule{0ex}{2.5ex} V(s_0)-V(s) \right]}\ . \rule[-3ex]{0ex}{1ex}
\end{equation}
Integrating over the relevant upstream points from the nearest
potential maximum to $s$ yields the total local ion density at $s$
\begin{equation}
\rho(s) = \int_{s_{\mbox{\tiny max}}}^{s} \rho_{s_0}(s) \, \md s_0\ .
\end{equation}
Using this method, and assuming that the residual gas density in the
linac sections is reduced by a factor 100 as specified in
Tab.~\ref{tb:erlpara}, we find the linear ion densities for the
Cornell ERL shown in Fig.~\ref{fg:ERL_lambda}.

The two sections of the ERL with the largest linear ion densities are
located at $s \approx 960$m (section A) and at $s \approx 1760$m
(section B).  We therefore studied these sections in detail by
performing numerical ion and beam simulations. The resulting
equilibrium ion densities can be found in Figs.~\ref{fg:ERL_B} and
\ref{fg:ERL_A}. The figures demonstrate that our analytical
approximation agrees well with the simulations even though the
approximation assumes that all ions are in the center of the electron
beam. We checked that the agreement becomes perfect in this case.

To reduce the impact of the two sections, additional clearing 
electrodes can be placed between the maximum and minimum of the potential.
We find that by including one extra electrode, the emittance increase can be
reduced from 1.04pm to 0.46pm for section A, and from 1.55pm to 0.71pm 
for section B.

\begin{figure}
\includegraphics[width=0.5\textwidth]{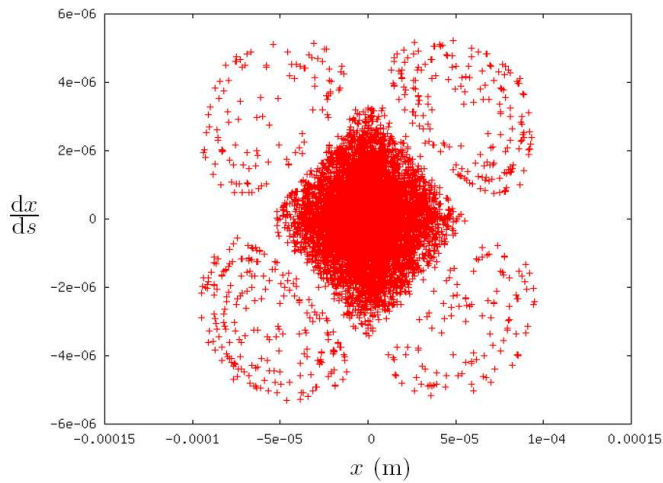}
\caption{Electron beam phase-space plot after transversing 
a 34m long region with medium high ion density 1000 times.}
\label{fg:islands}
\end{figure}

To get a better estimate of the ion effects in the full ERL, we also
simulated the ion distribution in a 34m long region around $s = 1530$m, 
which corresponds to one of the medium high peaks in
Fig.~\ref{fg:ERL_lambda}. Sending the beam through this section 1000
times results in the phase-space distribution in
Fig.~\ref{fg:islands}.
We find that after 1000 passes through the ion field, 6.2\% of the
beam electrons have left the main bunch and have migrated to the four
separated islands in phase space. The electronic version of this paper
contains an animation of this development of the beam halo.

\subsection{Outlook for non-round systems}

\begin{figure}
\includegraphics[width=0.5\textwidth]{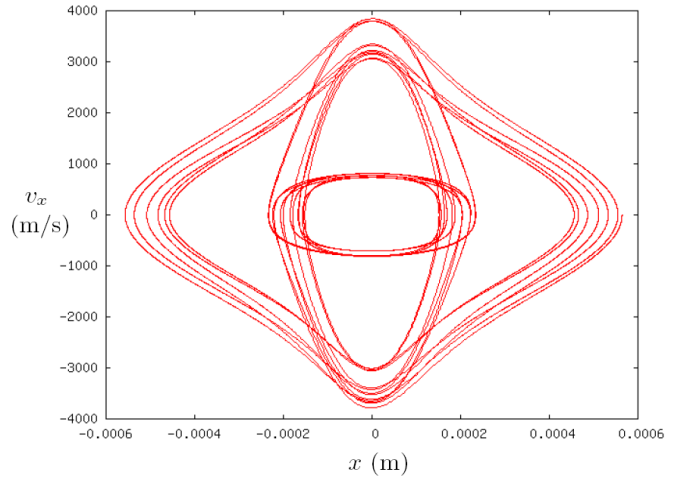}
\caption{Example of non-periodic transverse ion 
motion in a non-round beam setup.}
\label{fg:3dbeam}
\end{figure}

While we have argued that our circular beam model should be quite
accurate, it would be good to extend our simulations to non-round
beams. However, in that case the ion propagation cannot be simulated
in the same way using adiabatic invariants. Two independent invariant
quantities would be needed to account for oscillations in the $x$ and
$y$ directions. The problem is that the action integrals cannot be
defined as the area in a phase space plane, because the ion motion is
generally not a simple superposition of two independent periodic
oscillations in phase-space planes. An example of such phase-space
motion can be seen in Fig.~\ref{fg:3dbeam}.

\section{Conclusion}
We computed the ion distribution produced by an electron beam when
ion-clearing electrodes are installed. The transverse ion
distributions are shown to strongly peak in the beam's center,
producing very nonlinear forces on the electron beam. These can
produce strong perturbations of beam properties leading to emittance
growth and halo development. These simulations rely on fast algorithms
that take advantage of adiabatic invariants and scaling properties of
Maxwell's equations and the Lorentz force.

Our results are very relevant for high current Energy Recovery Linacs,
where ions are produced relatively quickly, and where clearing gaps in
the electron beam cannot easily be used for ion elimination. As an
example we used the Cornell Energy Recovery Linac project.  For
simplicity we only consider the case of a circular electron beam of
changing diameter. However, we parameterize this model to approximate
the non-round beams of the ERL well. We found suitable places for
clearing electrodes and computed the equilibrium ion density and the
electron-emittance growth it would produce even after installing extra
clearing electrodes at the two most dangerous places. We also showed
that the nonlinear-ion forces would lead to a significant halo
development.

\subsection*{Acknowledgments}
This work has been supported by NSF cooperative agreement PHY-0202078.

\end{document}